\documentclass[12pt]{iopart}
\usepackage{amssymb}
\usepackage{setstack}
\usepackage{verbatim}
\usepackage{bm}
\usepackage{graphicx}
\newcommand{\bq}{\begin{equation}}
\newcommand{\eq}{\end{equation}}
\newcommand{\bn}{\begin{eqnarray}}
\newcommand{\en}{\end{eqnarray}}

\begin{document}

\newcommand{\cc}{{\bf\Large C }}
\newcommand{\hide}[1]{}
\newcommand{\tbox}[1]{\mbox{\tiny #1}}
\newcommand{\half}{\mbox{\small $\frac{1}{2}$}}
\newcommand{\sinc}{\mbox{sinc}}
\newcommand{\const}{\mbox{const}}
\newcommand{\trc}{\mbox{trace}}
\newcommand{\intt}{\int\!\!\!\!\int }
\newcommand{\ointt}{\int\!\!\!\!\int\!\!\!\!\!\circ\ }
\newcommand{\eexp}{\mbox{e}^}
\newcommand{\EPS} {\mbox{\LARGE $\epsilon$}}
\newcommand{\ar}{\mathsf r}
\newcommand{\re}{\mbox{Re}}
\newcommand{\bmsf}[1]{\bm{\mathsf{#1}}}
\newcommand{\dd}[1]{\:\mbox{d}#1}
\newcommand{\abs}[1]{\left|#1\right|}
\newcommand{\ket}[1]{| #1 \rangle}
\newcommand{\bra}[1]{\langle #1 |}
\newcommand{\mbf}[1]{{\mathbf #1}}
%\definecolor{red}{rgb}{1,0.0,0.0}

\title{Initial-Phase Spectroscopy as a Control of Entangled Systems}
\author{Levente Horvath$^{1,2}$ and Zbigniew Ficek$^{3}$}
\eads{\mailto{zficek@kacst.edu.sa}}
\address{$^{1}$Department of Physics, School of Physical Sciences, The University of Queensland, Brisbane, QLD 4072, Australia}
\address{$^{2}$Centre for Quantum Computer Technology, Department of Physics, Macquarie University, Sydney, NSW 2109, Australia }
\address{$^{3}$The National Centre for Mathematics and Physics, KACST, P.O. Box 6086, Riyadh 11442, Saudi Arabia}

\date{today}

\begin{abstract}
We introduce the concept of initial-phase spectroscopy as a control of the dynamics of entangled states encoded into a two-atom system interacting with a broadband squeezed vacuum field. We illustrate our considerations by examining the transient spectrum of the field emitted by two systems, the small sample (Dicke) and the spatially extended (non-Dicke) models. It is found that the shape of the spectral components depends crucially on the relative phase between the initial entangled state and the squeezed field. We follow the temporal evolution of the spectrum and show that depending on the relative phase a hole burning can occur in one of the two spectral lines. We compare the transient behavior of the spectrum with the time evolution of the initial entanglement and find that the hole burning can be interpreted as a manifestation of the phenomenon of entanglement sudden death. In addition, we find that in the case of the non-Dicke model, the collective damping rate may act like an artificial tweezer that rotates the phase of the squeezed field.
\end{abstract}

\maketitle

\section{Introduction}

Recently there have been considerable interest in practical applications of squeezed light. The  interest stems from the recognition and experimental demonstrations that squeezed light could be used as a resource for entangling atomic ensembles~\cite{ly00,d05}, noise-free information transfer and processing of entanglement~\cite{ot06,nm06}. Testimony to this ability is provided by experimental observations of such effects as quantum teleportation~\cite{hp05,sk06}, quantum erasing~\cite{tb00,kk03}, quantum dense coding~\cite{mw05,wc06,ss09}, creation of entangled cluster states~\cite{s8,s9,s10,s11}, and entanglement storage in atomic ensembles~\cite{la04,cd08,bo09}. 

The early studies of squeezed light and its applications showed that atomic dynamics and fluorescence can be significantly modified in the presence a squeezed vacuum field~\cite{lk87}. The reason is in an unequal phase-dependent redistribution of the fluctuations (noise) between the two quadrature components of the field. It has been shown that this modification can be a source of many novel effects and techniques in quantum optics~\cite{pa93,fds,df04}. In particular, the interaction of atoms with a squeezed vacuum field can lead to the inhibition of atomic phase decay~\cite{ga86}, subnatural linewidths~\cite{pg87,pg88}, hole burning and dispersive profiles in the fluorescence and absorption spectra~\cite{sw94,zs97}, a linear dependence on the intensity of the two-photon absorption rate~\cite{gb89,jg90,fd91,fd93,gp95}, transparency and gain without population inversion~\cite{fb93,fs94}. The fluorescence and absorption spectra were found to exhibit a striking dependence on the relative phase of the squeezed vacuum and the coherent driving field~\cite{cl87,pa90,rz88}. 
Depending on the phase, the spectral lines can exhibit a subnatural or supernatural linewidth.

Although the phase dependent effects are usually predicted for systems that interact with a squeezed vacuum and are simultaneously driven by a coherent field, there can be interesting phase-dependent effects present even in the absence of the coherent field. This situation might be encountered when a system is initially prepared in a superposition state of its energy eigenstates. To put it another way, phase dependent effects are expected to vanish whenever the initial state of the system is an eigenstate of the energy, because then the emitted photon has no phase. Entangled states of a multi-atom system could provide a reference phase for the phase of the squeezed vacuum field. With the recent great interest in the dynamics of entanglement in atomic systems, it is worth to study phase dependent changes in the time evolution of entangled states of two atoms coupled to a squeezed vacuum field. Needless to say, the relative phase could be used as a control of the evolution of the entangled system.

In this paper we consider what may be termed "initial-phase spectroscopy" to search for a directly measurable signature of the abrupt loss of entanglement $-$ the phenomenon of sudden death of entanglement~\cite{zyc01,eb04,ft06,ey07,ye09,al07,xl10,mm09,zf10}. We study the transient spectrum of the field emitted from two two-level atoms initially prepared in a pure (phase dependent) entangled state and spontaneously decaying in a squeezed vacuum field. The squeezed field is treated as a correlated phase dependent noise environment. It is well known that an entangled state has a phase and nonvanishing correlations and therefore it is interesting to know how these initial correlations affect the spectrum when the system is subjected to a correlated phase dependent noise.
We illustrate our considerations by examining two different models of the two-atom system, in which phase dependent effects arise dramatically. These are the small sample (Dicke) model~\cite{Dic54} and the spatially extended (non-Dicke) model~\cite{le70,Fic02}. In both cases, the proposed initial-phase spectroscopy is found to be able to detect the phenomenon of entanglement sudden death by observing changes in the transient spectrum. These changes would also reveal other novel features that could provide a measure of the relative phase between two oscillating atomic dipoles. The transient spectra are produced in the following manner: in the time period before $t=0$ the atoms are prepared in an entangled state. At some time $(t=0)$ following the preparation of the atoms in the entangled state, a broadband squeezed vacuum field is applied to the atoms. Then, we trace the time evolution of the fluorescence spectrum from $t=0$ to the steady-state distribution by monitoring the spectrum as a function of time for different relative phases of the initial entangled state and the applied squeezed vacuum field. 

We interpret the results in terms of populations and two-photon coherence between collective states of the system and derive an analytical formula of the spectrum in the limit of the well separated spectral lines. It is well known that the intensity of a spectral line corresponding to transition between two collective states is proportional to the population of the state from which the transition originates. We find the intensity of one of the two spectral lines depends not only on the population of the upper state but also on the two-photon coherence. This dependence introduces the possibility to obtain the phase sensitive magnitude of this spectral line. In the course of the calculations, we observe that the sensitivity to the relative phase has an additional feature characteristic of entangled properties of the system. We find that the hole burning in the spectral line occurs at time when the atoms become disentangled. We are therefore led to the conclusion that observations of the hole burning in the spectral line provide a measurable signature of the phenomenon of entanglement sudden death. 

The paper is organized as follows. We start by introducing the system and the definition of the transient spectrum. In section~\ref{sec3}, we derive the equations of motion for the density matrix elements which we then solve for two different cases: the small sample (Dicke) model corresponding to atoms confined to a region of dimensions much smaller than the resonant wavelength, in section~\ref{sec4}, and for a spatially extended (non-Dicek) model where the atoms are confined to a region of dimensions comparable and larger than the resonant wavelength, in section~\ref{sec5}. Finally, in section~\ref{sec6}, we summarize our results.

\section{Transient fluorescence spectrum}\label{sec2}

We propose to study phase dependent effects of an entangled system in terms of the transient spectrum of the fluorescence field emitted by a two-atom system interacting with a broadband multi-mode squeezed vacuum field. Different definitions of the transient spectrum have been proposed, with their weaknesses and strengths summarized in~Ref.~\cite{Cre82}. As we shall be interested in signatures of the phenomenon of entanglement sudden death, we assume that the system starts at $t=0$ from a well defined initial entangled state and use the time dependent physical spectrum~\cite{ew77,ek80,ht82,hc85} 
\begin{eqnarray}
S(\omega,t) &=& 2\Gamma_{d}\sum_{i,j=1}^{2} \Gamma_{ij} \int_{0}^{t} {\rm d}t_{1} \int_{0}^{t} {\rm d}t_{2} \nonumber\\
&&\times {\rm e}^{-(\Gamma_{d}-i\omega)(t-t_{1})} {\rm e}^{-(\Gamma_{d} + i\omega)(t -t_{2})} \langle S^+_i(t_{1}) S^{-}_{j}(t_{2})\rangle ,\label{e1} 
\end{eqnarray}
where $t$ is the elapsed time after the system was prepared in the entangled state, $2\Gamma_{d}$ is the bandwidth of the detector which has its transmission peak at frequency~$\omega$, $S^+_i$ and $S^-_i$ are the dipole raising and lowering operators, respectively, of the~$i$th atom, $\Gamma_{ii}=\Gamma$ are the spontaneous emission decay rates of the atoms, equal to the Einstein $A$ coefficient for spontaneous emission, and $\Gamma_{ij}=\Gamma_{ji}, (i\neq j)$ is the collective damping which results from an exchange of photons between the atoms. The collective damping depends on the separation between the atom and the orientation of the atomic dipole moments with respect to the interatomic axis. The analytic expression for $\Gamma_{12}$ reads~\cite{le70,Fic02}
\begin{eqnarray}
\Gamma_{12} &=& \frac{3}{2}\Gamma\left\{ \left[1-\left( \hat{\mu}\cdot \hat{r}_{12}\right)^{2} \right] \frac{\sin\left( kr_{12}\right)}{kr_{12}}\right.  \nonumber \\
&+&\left. \left[ 1- 3\left( \hat{\mu}\cdot\hat{r}_{12}\right)^{2} \right]\left[ \frac{\cos\left(kr_{12}\right) }
{\left( kr_{12}\right) ^{2}} - \frac{\sin\left(kr_{12}\right) }{\left( kr_{12}\right)^{3}}\right]\right\} ,\label{5.9}
\end{eqnarray}
where $\hat{\mu}$ is the unit vector along the dipole moments of the atoms, which we have assumed to be parallel $(\hat{\mu}=\hat{\mu}_{1}=\hat{\mu}_{2})$,  $\hat{r}_{12}$ is the unit vector in the direction of $\vec{r}_{12}$, $k =\omega_{0}/c$, and $r_{12}$ is the distance between the atoms. The parameter $\Gamma_{12}$ is a damped oscillatory function of $r_{12}$, that varies from $\Gamma_{12}=\Gamma$ for $r_{12}\rightarrow 0$ to $\Gamma_{12}=0$ for $r_{12}\rightarrow \infty$. It can have positive as well as negative values. 

We note here that the transient spectrum (\ref{e1}) includes effects of a finite bandwidth of the detector. In experimental practice, this implies that our calculations are appropriate to observations involving a spectrometer (for example, a Fabry-Perot interferometer) that analyses the frequency spectrum of the emitted field.

The two-time correlation function, appearing in Eq.~(\ref{e1}), is found from the solution of the master equation for the density operator of the system and quantum regression theorem. We assume that the system we consider is composed of two identical two-level atoms (qubits) interacting with a broadband reservoir whose modes are in a squeezed vacuum state. We study the behavior of the system in terms of the density operator of the atoms, which in the Born-Markov and the rotating-wave approximations satisfies the following master equation
\begin{eqnarray}
\label{eq:master}
\dot{\rho} =\frac{1}{i\hbar}\left[H_{\rm eff},\rho\right]+ {\cal L}_{\rm sp} \rho ,
\end{eqnarray}
where we have explicitly distinguished two evolution terms, a coherent evolution under 
the non-Hermitian Hamiltonian
\begin{eqnarray}\label{eq:Heff}
H_{\rm eff} &=& \hbar \sum_{i=1}^2 \left(\omega_{0} S_i^z +
\sum_{j\ne i} \Omega_{ij}S_i^+S_j^{-}\right) \nonumber \\
&-&\frac{1}{2}i\hbar \sum_{i,j=1}^2 \Gamma_{ij}\left\{\left(1+ N\right)S_i^+S_j^- 
+ N S_i^- S_j^{+}\right. \nonumber\\
&-&\left. \left(M S_i^+ S_j^{+}{\rm e}^{-2i\omega_{s}t} + M^* S_i^- S_j^{-}{\rm e}^{2i\omega_{s}t}\right)\right\} ,
\end{eqnarray}
and an incoherent evolution that is solely due to spontaneous emission events
\begin{eqnarray}
\label{eq:rhospon}
{\cal L}_{\rm sp}\rho &= & \sum_{i,j=1}^{2}\Gamma_{ij}\left[\left(1 + N \right)
S_i^-\rho S_j^+  + N S_i^+\rho S_j^- \right]\nonumber \\
&-&\sum_{i,j=1}^{2}\Gamma_{ij} \left(M S_i^+\rho S_j^{+}{\rm e}^{-2i\omega_{s}t}  
+ M^*S_i^-\rho S_j^{-}{\rm e}^{2i\omega_{s}t}\right) .
\end{eqnarray}
Here $M=\vert M\vert\exp(i\phi_s)$ is the magnitude of two-photon correlations within the squeezed vacuum field,~$\phi_s$ is its phase, $N$ describes the number of photons in 
the field, $\omega_{0}$ is the atomic transition frequency, and $\omega_{s}$ is the carrier frequency of the squeezed field that we shall assume to be equal to the atomic transition frequency. The parameter $\Omega_{12}=\Omega_{21}$ stands for the dipole-dipole interaction strength between the atoms~\cite{le70,Fic02}.

The first line in the effective Hamiltonian (\ref{eq:Heff}) describes the coherent evolution between the atoms due to dipole-dipole coupling, the second line describes incoherent damping of the atoms due to the presence of the broadband squeezed vacuum, and the third describes the coherent excitation of the atoms by the squeezed field. The term ${\cal L}_{\rm sp}\rho$ appears as an another source of incoherent damping.

\begin{figure}[h]
\begin{center}
\resizebox*{7cm}{!}{\includegraphics{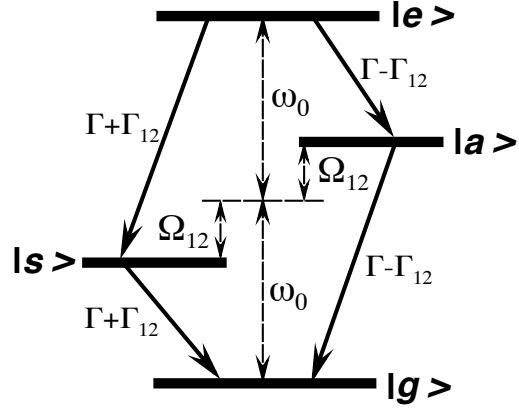}}
\caption{The energy-level diagram of a two-atom system with the allowed spontaneous transitions occurring with rates $\Gamma +\Gamma_{12}$ and $\Gamma -\Gamma_{12}$.}
\label{fig1}
\end{center}
\end{figure}

It is convenient to study the atomic dynamics in terms of the collective states of the system, which are the eigenstates of the Hamiltonian
\begin{eqnarray}
H_{c} = \hbar \sum_{i=1}^2 \omega_{0} S_i^z + \hbar\sum_{i\ne j=1}^2
 \Omega_{ij} S_i^+S_j^- . \label{e6a}
\end{eqnarray}
Here the dipole-dipole interaction can be viewed as a mixer of the atomic bare states. It produces collective states that are found by the diagonalization of the Hamiltonian~(\ref{e6a}):
\begin{eqnarray}
\ket e &=&  \vert e_1 \rangle \vert e_2\rangle, \quad 
\ket g = \vert g_1 \rangle \vert g_2\rangle ,\nonumber\\
\ket s &=& \frac{1}{\sqrt{2}} \left(\vert e_1\rangle\vert g_2\rangle 
+\vert g_1\rangle\vert e_2\rangle\right)  ,\nonumber\\
\ket a &=& \frac{1}{\sqrt{2}} \left(\vert e_1\rangle\vert g_2\rangle 
-\vert g_1\rangle\vert e_2\rangle\right) ,\label{e6}
\end{eqnarray}
where $\vert e_i\rangle$ and~$\vert g_i\rangle $ represent the excited and ground state of the $i$th atom, respectively. In the basis of the collective states, the system is composed of four non-degenerate energy states, shown in Fig.~\ref{fig1}, the ground state $\ket g$ of energy $E_{g}=0$, two intermediate states $\ket s$ and $\ket a$ of energies $E_{s,a}= \hbar (\omega_{0}\pm \Omega_{12})$, and the upper state $\ket e$ of energy $E_{e}=2\hbar\omega_{0}$. The existence of the maximally entangled states, the symmetric $\ket s$ and anti-symmetric $\ket a$ states, reflects the presence of the coherent dipole-dipole interaction between the atoms. 

Since atoms interact with each other, the collective state picture provides a good approach for studying the initial phase problem. 
In what follows, we use the collective states as the basis for the density operator of the system, and find that the two-time correlation function appearing in Eq.~(\ref{e1}) can be written in terms of the density matrix elements as
\begin{eqnarray}
\sum_{i,j=1}^2 \Gamma_{ij}  \langle S^+_i(t) S^-_j(t+\tau)\rangle  &=& \left(\Gamma+\Gamma_{12}\right) \left[ \rho_{es}(t + \tau) + \rho_{sg}(t + \tau)\right] \nonumber \\
&+& \left(\Gamma-\Gamma_{12}\right) \left[\rho_{ag}(t + \tau) - \rho_{ea}(t + \tau)\right] ,\label{eq:S-S+}
\end{eqnarray}
with the initial $\tau=0$ values determined by the one-time density matrix elements
\begin{eqnarray}
\label{eq:SpecIni}
\rho_{es}(t+\tau)|_{\tau=0} &\rightarrow \rho_{ee}(t) ,  \ \rho_{sg}(t+\tau)|_{\tau=0}\rightarrow \rho_{ss}(t) , \nonumber \\
\rho_{ea}(t+\tau)|_{\tau=0} &\rightarrow -\rho_{ee}(t) ,  \ \rho_{ag}(t+\tau)|_{\tau=0}\rightarrow \rho_{aa}(t) ,\nonumber\\
\rho_{gs}(t+\tau)|_{\tau=0}&\rightarrow \rho_{ge}(t) ,\ \rho_{ga}(t+\tau)|_{\tau=0}\rightarrow -\rho_{ge}(t) .
\end{eqnarray}
One can see that the correlation function appears as a sum of the coherences which cascade down the atomic levels. Note that there are no correlations in the transitions between and inside the cascade down channels $\ket e \rightarrow \ket s \rightarrow \ket g$ and $\ket e \rightarrow \ket a \rightarrow \ket g$. This property allows as to analyze spectral properties of each of the transitions separately. 

To introduce initial-phase spectroscopy, we first consider a simplified two-atom system, namely the Dicke model, in which the transition channel $\ket e\rightarrow\ket a\rightarrow\ket g$ does not participate in the dynamics of the system. As we shall see, this will allow us to describe and characterise initial-phase spectroscopy using analytic methods. What do we mean by "initial-phase" spectroscopy? A squeezed vacuum itself depends on the phase $\phi_{s}$, but no phase dependent effects are observed until a reference phase is introduced. Usually, the reference phase is provided by a coherent laser field. 

An another way to introduce the reference phase is to prepare the system in an entangled state and study the dependence of the evolution on the relative phase~$\Delta\phi$ between the initial entangled state and the squeezed vacuum. This is what we call the "initial-phase spectroscopy" that could allow us to study the dependence of the decoherence process on the phase and to identify entangled states that could be prevented from decoherence. We look for signatures of slowed down or accelerated decoherence in the transient spectrum of the emitted fluorescence~field. 

Usually, decoherence processes in a given system are studied by analyzing widths of the spectral lines. They are determined by asymptotic, exponentially decaying correlation functions whose Fourier transforms give spectral (Lorentzian) lines. A narrowing of the spectral lines below the natural linewidth is interpreted as a stabilization of the quantum fluctuations and hence the reduction of decoherence. However, the recent studies on entanglement have revealed that some entangled states can decay is a non-exponential way leading to the phenomenon of entanglement sudden death~\cite{eb04,ey07,ye09,zf10}. This kind of decay is not reflected in the spectral linewidths. As we shall see, the non-exponential decay can be reflected in magnitudes of the spectral~lines.

\section{Evolution of the density matrix elements}\label{sec3}

The spectrum is determined by the coherence terms that govern spontaneous and stimulated transitions in the system. The coherences are solutions of the equations of motion that are found from the master equation~(\ref{e1}). In the rotating frame oscillating with the frequency of the squeezed field $\omega_{s}$, the equations of motion for the coherences responsible for transitions in the channel involving the symmetric state are in the form
\begin{eqnarray}
\dot{\tilde \rho}_{es} &=& -\left\{\frac{1}{2}\Gamma\left[n(2+a)+1\right]- i\Omega_{12}\right\}\tilde\rho_{es} \nonumber \\ 
&&+ \frac{1}{2}\Gamma (n-1)(1+a)\tilde\rho_{sg} + M\Gamma\left[a\tilde\rho_{gs} -\left(1+a\right)\tilde\rho_{se}\right]  , \nonumber\\
\dot{\tilde\rho}_{sg} &=& -\left\{\frac{1}{2}\Gamma\left[n(2+a)-1\right]+ i\Omega_{12}\right\}\tilde\rho_{sg} \nonumber \\ 
&&+ \frac{1}{2}\Gamma\left(n+1\right)\left(1+a\right)\tilde\rho_{es}+ M\Gamma\left[a\tilde\rho_{se} -\left(1+a\right)\tilde\rho_{gs}\right] ,\label{e11}
\end{eqnarray}
and the equations of motion for the coherences responsible for transitions in the channel involving the antisymmetric state are
\begin{eqnarray}
\dot{\tilde\rho}_{ea} &=& -\left\{\frac{1}{2}\Gamma\left[n(2-a)+1\right] + i\Omega_{12}\right\}\tilde\rho_{ea} \nonumber \\ 
&&- \frac{1}{2}\Gamma (n-1)(1-a)\tilde\rho_{ag} + M\Gamma\left[a\tilde\rho_{ga} -\left(1-a\right)\tilde\rho_{ae}\right]   ,\nonumber \\
\dot{\tilde \rho}_{ag} &=& -\left\{\frac{1}{2}\Gamma\left[n(2-a)-1\right]- i\Omega_{12}\right\}\tilde \rho_{ag} \nonumber \\ 
&&+ \frac{1}{2}\Gamma \left(n+1\right)\left(1-a\right)\tilde\rho_{ea}+ M\Gamma\left[a\tilde\rho_{ae} -\left(1-a\right)\tilde\rho_{ga}\right] ,\label{e12}
\end{eqnarray}
where $n= (2N+1)$ and $a=\Gamma_{12}/\Gamma$ is a dimensionless collective damping parameter.  
 
The coherences oscillate with frequencies shifted from $\omega_{0}$ by the amount $\Omega_{12}$, the dipole-dipole interaction strength, and the squeezed field couples the coherences inside the channels. In addition, the squeezed correlation $M$ couples the coherences to their conjugates. It is interesting to note that there is no coupling between the symmetric and antisymmetric channels. 

The solution of Eqs.~(\ref{e11}) and (\ref{e12}) depends, of course, on the initial state of the system. We shall assume in all our considerations that the initial state of the atomic system is  the maximally entangled Bell state
\begin{eqnarray}
\vert \psi (t = 0)\rangle =\frac{1}{\sqrt{2}}\left(\ket g  + {\rm e}^{i\phi_B}\ket e \right) ,\label{e13}
\end{eqnarray}
which corresponds to an excitation of the two-atom system into a coherent superposition of the states $\ket g$ and $\ket e$. Here, $\phi_{B}$ is the (fixed) phase of the state that serves as the reference phase for the squeezed field. The application of the phase dependent initial state allows the phase control of the fluorescence spectrum and the evolution of the initial entanglement. 

The initial non-zero density matrix elements of the system prepared in the entangled 
state~(\ref{e13}), before the application of the squeezed field, are
\begin{eqnarray}
\rho_{ee} (0) &=& \rho_{gg}(0)=  \frac{1}{2} ,\nonumber\\  
\rho_{eg}(0) &=& \frac{1}{2}{\rm e}^{i\phi_B} ,\ \rho_{ge}(0)=\frac{1}{2}{\rm e}^{-i\phi_B} ,\label{e14}
\end{eqnarray}
and all the remaining density matrix elements are zero. 

Our task is to trace the time evolution of the spectrum from the initial state~(\ref{e13}) to the steady-state distribution. The evaluation of the transient spectrum requires the knowledge of the time evolution of the populations of the collective states and the two-photon coherence between them. These are determined from the following equations of motion
\begin{eqnarray}
\dot{\rho}_{ee} &=& -(n+1)\Gamma\rho_{ee} +2\vert M\vert\Gamma_{12} \rho_u \nonumber \\
&&+ \frac{1}{2}(n-1)\Gamma\left[\left(1+a\right)\rho_{ss} + \left(1-a\right)\rho_{aa}\right] ,\nonumber \\ 
\dot{\rho}_{ss} &=& \frac{1}{2}\left(1+a\right)\Gamma\left[(n-1) - (3n - 1)\rho_{ss}\right. \nonumber\\
&&- \left.  (n-1) \rho_{aa} + 2\rho_{ee} - 2\vert M\vert \rho_u \right] ,\nonumber \\
\dot{\rho}_{aa} &=& \frac{1}{2}\left(1-a\right)\Gamma\left[(n-1)\!-\!(3n - 1)\rho_{aa}\right. \nonumber\\
&&- \left. (n-1) \rho_{ss} + 2\rho_{ee} + 2\vert M\vert \rho_u \right] ,\nonumber \\
\dot{\rho}_u &=& \Gamma_{12}\vert M\vert - n\Gamma \rho_u \nonumber \\
&&- \vert M\vert\Gamma\left[ \left(1+2a\right)\rho_{ss} 
-\left(1-2a\right)\rho_{aa}\right] ,\label{e15}
\end{eqnarray}
where~$\rho_u=(\rho_{eg}\exp(-{\rm i}\phi_s)+\rho_{ge}\exp({\rm i}\phi_s))/2$ is the real part of the two-photon coherence. 

Note that the initial value of $\rho_{u}(0) = \cos(\Delta\phi)/2$ depends on the relative phase $\Delta\phi =\phi_{B}-\phi_{s}$ between the initial entangled state and the input squeezed field. This is the quantity that brings the phase dependent features into the time evolution of the fluorescence field and its spectral distribution. It is also interesting to note that the evolution of the populations is insensitive to the dipole-dipole coupling strength~$\Omega_{12}$. This is due to the broadband nature of the applied squeezed field, which makes the squeezing parameters $N$ and $M$ independent of frequency.

\section{Dicke Model}\label{sec4}

We first concentrate on a simplified model of the two-atom system, the Dicke model, where the evolution of the system through the antisymmetric channel is ignored. Physically, it corresponds to a system of two atoms confined to a region which has linear dimensions much smaller than the resonant wavelength, $r_{12}\ll \lambda_{0}=2\pi c/\omega_{0}$, so that both atoms experience the same amplitude and phase of the field. In other words, the atoms act collectively with a common dipole phase. Mathematically, the model is described by the same set of equations of motion as Eqs.~(\ref{e11}), (\ref{e12}) and (\ref{e15}), but with~$\Gamma_{12}$ replaced by~$\Gamma$. Effectively, the model behaves as a three-level cascade system with upper state $\ket e$, intermediate state $\ket s$ and ground state $\ket g$. The transitions in the cascade are both damped with the rate~$2\Gamma$. 
\begin{figure}[h]
\resizebox*{8.5cm}{!}{\includegraphics{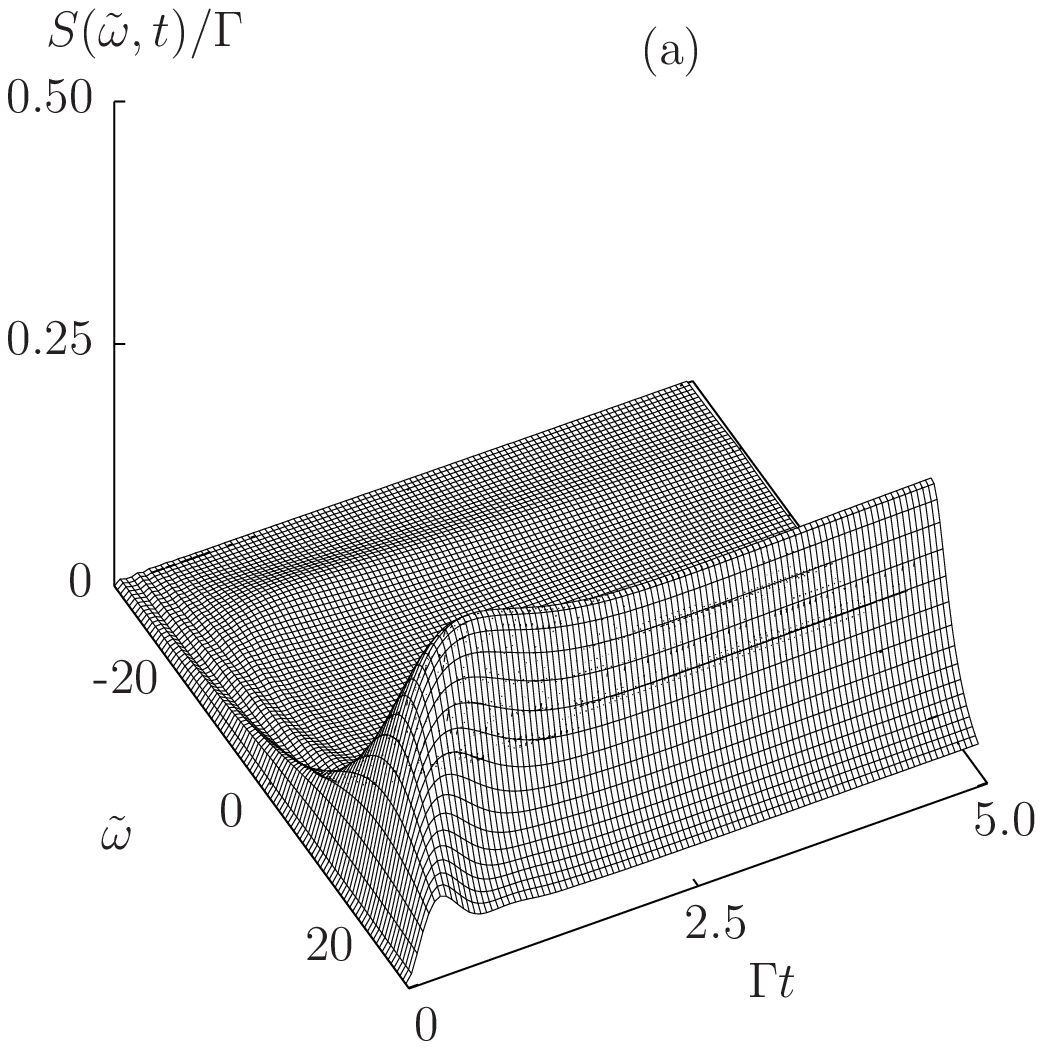}}
\resizebox*{7cm}{!}{\includegraphics{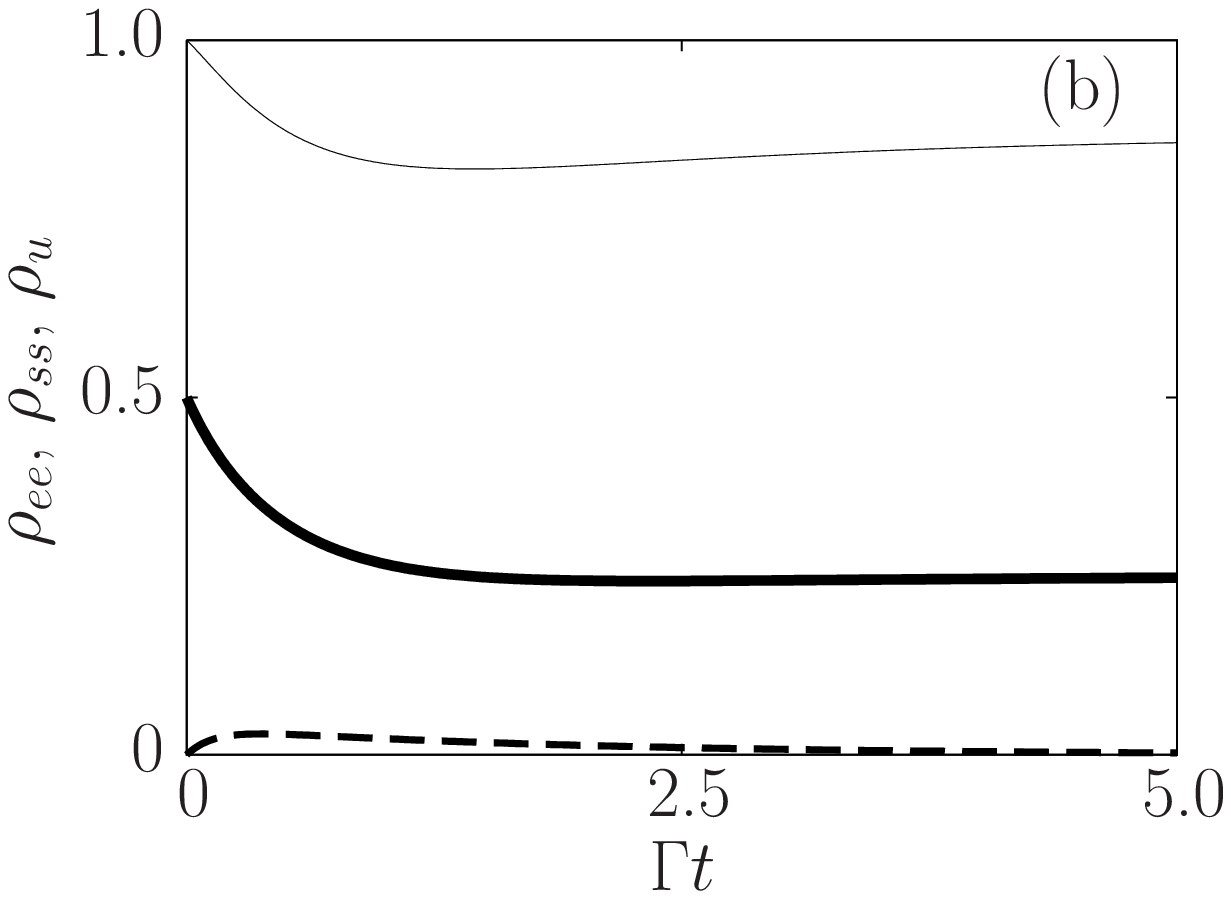}}
\caption{(a) The transient spectrum $S(\tilde{\omega},t)$ as a function of the normalized frequency~$\tilde{\omega} = (\omega -\omega_{0})/\Gamma$ and time $\Gamma t$ for the Dicke model with $\Gamma_{d}=2\Gamma$, $\Delta\phi=0$, $N=0.5$, $\Omega_{12}=20\Gamma$, and $\vert M\vert=\sqrt{N (N+1)}$. The frame (b) illustrates the time evolution of the populations~$\rho_{ee}$ (thick line), and $\rho_{ss}$ (thick dashed line) together with the two-photon coherence $\rho_u$ (thin line). The atoms were initially in the Bell state~(\ref{e13}).}
\label{fig2}
\end{figure}

In what follows, we calculate the transient fluorescence spectrum by solving the equations of motion for the density matrix elements with $\Gamma_{12}=\Gamma$, from which we then  evaluate the spectrum~(\ref{e1}) in the the limit of well separated spectral lines, $\Omega_{12}\gg \Gamma N$. 

Figure~\ref{fig2}(a) shows the transient spectrum calculated for the relative phase $\Delta\phi=0$. The system was initially prepared in the Bell state (\ref{e13}). We see that the amplitude of the peak at $\omega =\omega_{0}-\Omega_{12}$ is practically zero for all times, whereas the amplitude of the other peak is large. The amplitude of the peak at $\omega=\omega_{0}+\Omega_{12}$ increases in time with the rate~$\Gamma$, it passes through a maximum and then slowly decays towards a stationary value. A close look at the transient evolution of the amplitude reveals the existence of a minimum (saddle point) that appears at a time near $\Gamma t=1$. The existence of the saddle point can be interpreted as a result of switching the atomic dynamics from that governed by the initial correlation to that governed by the correlation transferred to the atoms from the squeezed field. 

In order to interpret the numerical results, we calculate analytically the transient spectrum in the limit of $\Omega_{12}\gg \Gamma N$ corresponding to the case of well separated spectral components. The spectrum depends on the bandwidth of the detector, for simplicity and without loss of generality we will assume that $\Gamma_{d}\rightarrow \infty$. The limit of $\Gamma_{d}\rightarrow \infty$ corresponds to a very broad detector. In this case, the spectrum (\ref{e1}) reduces to a finite Fourier transformation of the two-time atomic correlation
\begin{eqnarray}
S(\omega,t) = {\rm Re} \sum_{i,j=1}^{2} \Gamma_{ij}\int_{0}^{t} {\rm d}\tau\, {\rm e}^{-i\omega\tau} \langle S^+_i(t) S^-_j(t + \tau)\rangle .\label{ee1} 
\end{eqnarray}
The spectrum might be interpreted as the rate of production of photons of frequency~$\omega$ at time $t$. 

Using Eqs.~(\ref{e11}) and (\ref{e12}), and neglecting all terms of order $\Omega_{12}^{-1}$ and smaller, we obtain the following expression for the spectrum
\begin{eqnarray}
S(\omega,t) & = & 2\Gamma \left\{
\frac{2\left(n+1\right)\rho_{ee}(t) -2|M|\rho_u(t)}
{4n^{2}+(\tilde{\omega} -\tilde{\Omega}_{12})^{2}}
+\frac{(n-1)\rho_{ee}(t)+|M|\rho_u(t)}
{n^{2}+ (\tilde{\omega} -\tilde{\Omega}_{12})^{2}}\right.\nonumber \\
&+& \left. \rho_{ss}(t)\left[ \frac{2(n-1)}{4n^{2} + (\tilde{\omega} +\tilde{\Omega}_{12})^{2}}
+ \frac{n+1}{n^{2} + (\tilde{\omega} +\tilde{\Omega}_{12})^{2}}\right]\right\} .\label{e16}
\end{eqnarray}
where $|M|=\sqrt{n^{2}-1}, \tilde{\omega} =(\omega -\omega_{0})/\Gamma$ and $\tilde{\Omega}_{12}= \Omega_{12}/\Gamma$.
The spectrum has a simple structure. In general, at a given time, the spectrum is a sum of four Lorentzian lines, with two overlapping lines at each of the transitions. The amplitude of the lines at the transition frequency $\omega =\omega_{0}-\Omega_{12}$ is proportional to the population $\rho_{ss}(t)$ of the symmetric state, whereas the amplitude of the lines at $\omega=\omega_{0}+\Omega_{12}$ is determined by the population $\rho_{ee}(t)$ of the upper state. However, most important from the point of view of this paper is that there is an additional contribution to the amplitude of the line at $\omega =\omega_{0}+\Omega_{12}$ from the coherence $\rho_{eg}(t)$. Evidently, the two-photon coherence and in particular its dependence on the relative phase can directly affect spontaneous transitions originating from the state $\ket e$.

It is clear from (\ref{e16}) that the absence of the peak at $\omega=\omega_{0}-\Omega_{12}$ results from the absence of the population in the symmetric state. This is also seen from Fig.~\ref{fig2}(b), where we plot the time evolution of the populations $\rho_{ee}(t), \rho_{ss}(t)$ and the coherence $\rho_{u}(t)$. The absence of the population in $\ket s$ is readily understood if we recall that there are two pathways for excitation of state $\ket e$. First, there is a stepwise excitation in which the system goes from $\ket g$ to $\ket s$ and then from $\ket s$ to $\ket e$. Second, there is a direct transition from $\ket g$ to $\ket e$ involving a simultaneous absorption of two photons. The two-photon correlations, the initial and that contained in the squeezed field, cause simultaneous absorption of two photons which results in the direct transfer of population from the ground state $\ket g$ to the upper state $\ket e$ without populating the intermediate state~$\ket s$. Thus, the buildup of the peak at $\omega=\omega_{0}-\Omega_{12}$ is suppressed.
\begin{figure}[h]
\resizebox*{8.5cm}{!}{\includegraphics{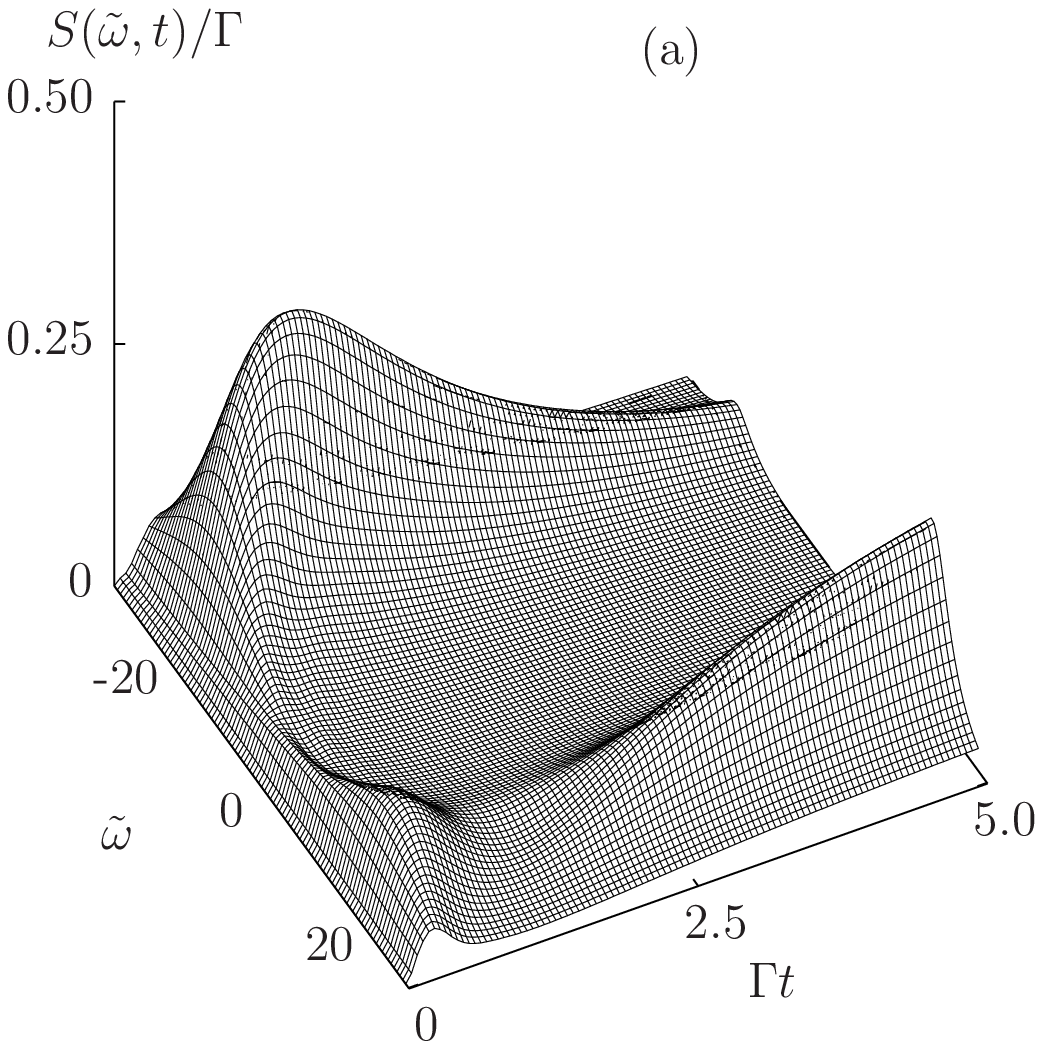}}
\resizebox*{7cm}{!}{\includegraphics{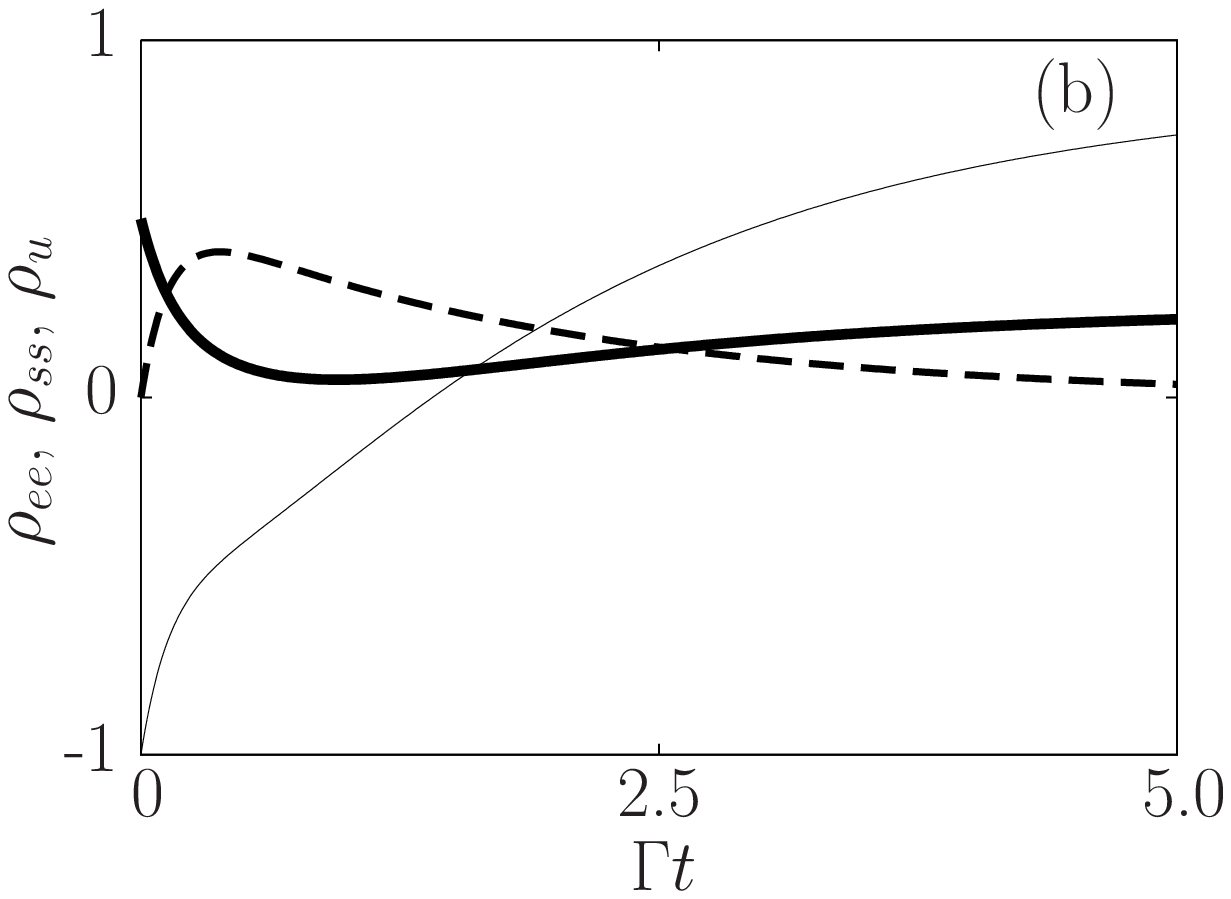}}
\caption{Same as in Fig.~\ref{fig2}, but for the relative phase $\Delta\phi=\pi$.}
\label{fig3}
\end{figure}

The dependence on time of the spectral features changes qualitatively when $\Delta\phi$ is varied. A change of $\Delta\phi$ from $0$ to $\pi$ leads to a hole burning in the higher frequency peak. This is shown in Fig.~\ref{fig3}, where we plot the spectrum for the same parameters as in Fig.~\ref{fig2} but $\Delta\phi=\pi$. We observe a rapid collapse of the amplitude of the peak at $\omega=\omega_{0}+\Omega_{12}$, which then evolves through a hole burning to a narrow peak as time varies from $t=0$ to $t=2\Gamma$. To explain the origin of the hole burning, we illustrate in frame (b) the time evolution of the populations $\rho_{ee}(t), \rho_{ss}(t)$ and the coherence $\rho_{u}(t)$. Evidently, at time when the hole is burned in the spectrum, the population $\rho_{ee}(t)$ is almost zero and $\rho_{u}(t)$ is negative. 

The hole burning, as illustrated in Fig.~\ref{fig3}(a) and its persistence for only small period of time around $\Gamma t=1$ can be understood as the result of the destructive interference between the initial and squeezing correlations that oscillate out-of phase when $\Delta\phi=\pi$~\cite{gp97}. An alternative explanation is as follows. The spectral line at $\omega =\omega_{0}+\Omega_{12}$, as seen from Eq.~(\ref{e16}) is composed of two Lorentzians of different widths and weights. Each of the weights can be positive or negative depending on the sign of the coherence $\rho_{u}(t)$. For $\Delta\phi =0$, the coherence $\rho_{u}(t)$ is positive for all time and much larger than the population $\rho_{ee}(t)$. In this case, the weight of the broader spectral component with the width, $2n\Gamma$ is negative while the weight of the narrower spectral component with the width, $n\Gamma$, is positive. Mollow~\cite{mo69} and Rice and Carmichael~\cite{rc88} pointed out that the negative weight of the broader Lorentzian can give rise to a spectrum with subnatural linewidth. Thus, the subtraction of the broader Lorentzian from the narrower one results in a narrowing of the spectral line and the reduction of the intensity of the line. The narrowing of the spectral line is not well visible in Fig. 2, but the reduction of the amplitude, a saddle point near $\Gamma t=1$ is well visible in the figure. The situation changes dramatically when one changes the relative phase to $\Delta\phi =\pi$. In this case, $\rho_{u}(t)$ is negative for short times so that the narrower Lorentzian can now have a negative weight while the broader Lorentzian has a positive weight. Thus, for times close to the value which minimizes the weight of the narrower Lorentzian, we obtain a hole burned in the spectrum~\cite{rc88}. 

The enhancement of the spectral peak at $\omega =\omega_{0}-\Omega_{12}$ reminds us the phenomenon of the Dicke superradiant pulse formulation~\cite{Cle03,Cle04}. Over a short time period, the population is transferred to the symmetric (superradiant) state and is then emitted with the enhanced, superradiant rate $\Gamma +\Gamma_{12}$, which is the damping rate of the population of the symmetric state. 

The hole burning can be connected with the phenomenon of entanglement sudden death~\cite{eb04,ey07,ye09,zf10}. To show this, we compare the time evolution of the spectrum with the evolution of an initial entanglement between the atoms. We quantify entanglement by calculating the concurrence, which can be evaluated explicitly from the density operator of the system as~\cite{woo}
\begin{eqnarray}
{\cal C} = \max\left(0,\sqrt{\lambda_{1}}-\sqrt{\lambda_{2}}-\sqrt{\lambda_{3}} -\sqrt{\lambda_{4}}\,\right) ,\label{4.4}
\end{eqnarray}
where the quantities $\lambda_{i}$ are the eigenvalues in decreasing order of the matrix
\begin{eqnarray}
  R=\rho\left(\sigma_{y}\otimes\sigma_{y}\right)\rho^{\ast}\left(\sigma_{y}\otimes\sigma_{y}\right) ,\label{4.5}
\end{eqnarray}
where $\rho^{\ast}$ denotes the complex conjugate of $\rho$ and $\sigma_{y}$ is the Pauli matrix. Concurrence varies from ${\cal C} =0$ for separable qubits to~${\cal C} =1$ for maximally entangled qubits, and the intermediate cases $0< {\cal C}<1$ characterize a partly entangled qubits. 

The concurrence is specified by the density matrix of a given system and thus is evaluated from the knowledge of the density matrix elements. For the system considered here of two atoms interacting with a squeezed vacuum field, the density matrix written in the basis of the collective states (\ref{e6}) has the $X$-state form
\begin{eqnarray}
  \rho = \left(
    \begin{array}{cccc}
      \rho_{gg} & 0 & 0 & \rho_{ge}  \\
      0 & \rho_{ss} & 0 & 0\\
      0 & 0 & \rho_{aa} & 0\\
      \rho_{eg}  & 0 & 0 &\rho_{ee}
    \end{array}\right) .\label{4.19}
\end{eqnarray}
This simple form of the density matrix leads to the concurrence of the form 
\begin{eqnarray}
  {\cal C}(t) = \max\left\{0,\,{\cal C}_{1}(t),\,{\cal C}_{2}(t)\right\} ,\label{4.11} 
\end{eqnarray}
where
\begin{eqnarray}
{\cal C}_{1}(t) = 2|\rho_{eg}(t)| -\left[\rho_{ss}(t)+ \rho_{aa}(t)\right] ,\label{4.12}
\end{eqnarray}
and
\begin{eqnarray}
{\cal C}_{2}(t) = |\rho_{ss}(t)-\rho_{aa}(t)| - 2\sqrt{\rho_{gg}\rho_{ee}} .\label{4.13} 
\end{eqnarray}
It is clear that the concurrence ${\cal C}(t)$ can always be regarded as being made up of the sum of nonnegative contributions ${\cal C}_{1}(t)$ and ${\cal C}_{2}(t)$ that are associated with
two different sources of entanglement in a two atom system. The contribution ${\cal C}_{1}(t)$ provides a measure of an entanglement produced by the two-photon coherence $\rho_{eg}$, whereas ${\cal C}_{2}(t)$ provides a measure of an entanglement produced by the populations of the maximally entangled states $\ket s$ and $\ket a$. 

It is easily verified, that with the initial Bell state~(\ref{e13}), ${\cal C}_{2}(t)$ is always negative, so that a possible entanglement in the system is solely determined by ${\cal C}_{1}(t)$.
\begin{figure}[h]
\resizebox*{8cm}{!}{\includegraphics{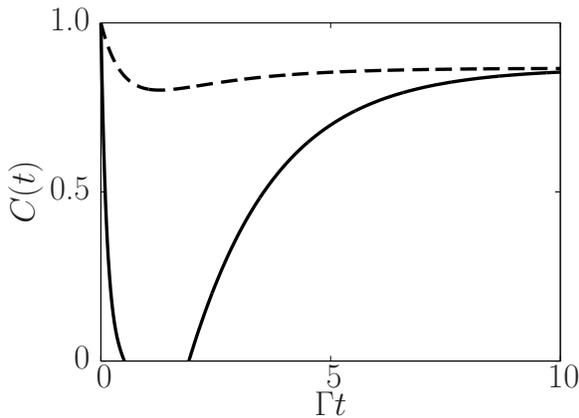}}
\caption{Time evolution of the concurrence for the Dicke model coupled to a broadband squeezed vacuum field with $N=0.5$, $\vert M\vert=\sqrt{N (N+1)}$, and for different relative phases: $\Delta\phi=\pi$ (solid line), $\Delta\phi =0$ (dashed line).}
\label{fig4}
\end{figure}

Figure~\ref{fig4} shows the time evolution of the concurrence for the same choice of parameters as in Figs. 2 and 3. We see a strong variation in the time evolution of the concurrence with the relative phase $\Delta\phi$. For the case of $\Delta\phi=0$, the atoms are entangled for all times with the initial entanglement decaying slowly towards a non-zero steady state value. The evolution of the initial entanglement is completely different when the phase is changed to $\Delta\phi =\pi$. In this case, the initial entanglement decays in a non-exponential fashion and disappears at a very short time of the evolution. Beyond that time, the atoms remain disentangled for a finite period of time, and then the entanglement reappears again~\cite{ft06,xl10}, and evolves asymptotically to a steady-state value that is independent of the phase $\Delta\phi$. A close look at the spectrum and the concurrence reveals that the entanglement disappears for the period of time when the hole occurs in the system. 
Thus, the hole burning in the spectrum can be regarded as a signature of the phenomenon of sudden death of entanglement. Hence, observation of the hole burning in the spectrum would be a striking confirmation of the entanglement sudden death.

\section{Non-Dicke Model}\label{sec5}

We now turn to the problem of phase effects in an extended atomic system. We relax the small sample approximation and consider dynamics of an extended atomic system, the so-called non-Dicke model, where we explicitly take into account a non-zero spatial separation between the atoms. The separation will act as a phase difference between the oscillating dipoles. As we shall see, this will introduce an additional phase to the dynamics of the system. In other words, the transient spectrum depends strongly on whether $\Gamma_{12}>0$ or~$\Gamma_{12}<0$. 
\begin{figure}[h]
\resizebox*{8.5cm}{!}{\includegraphics{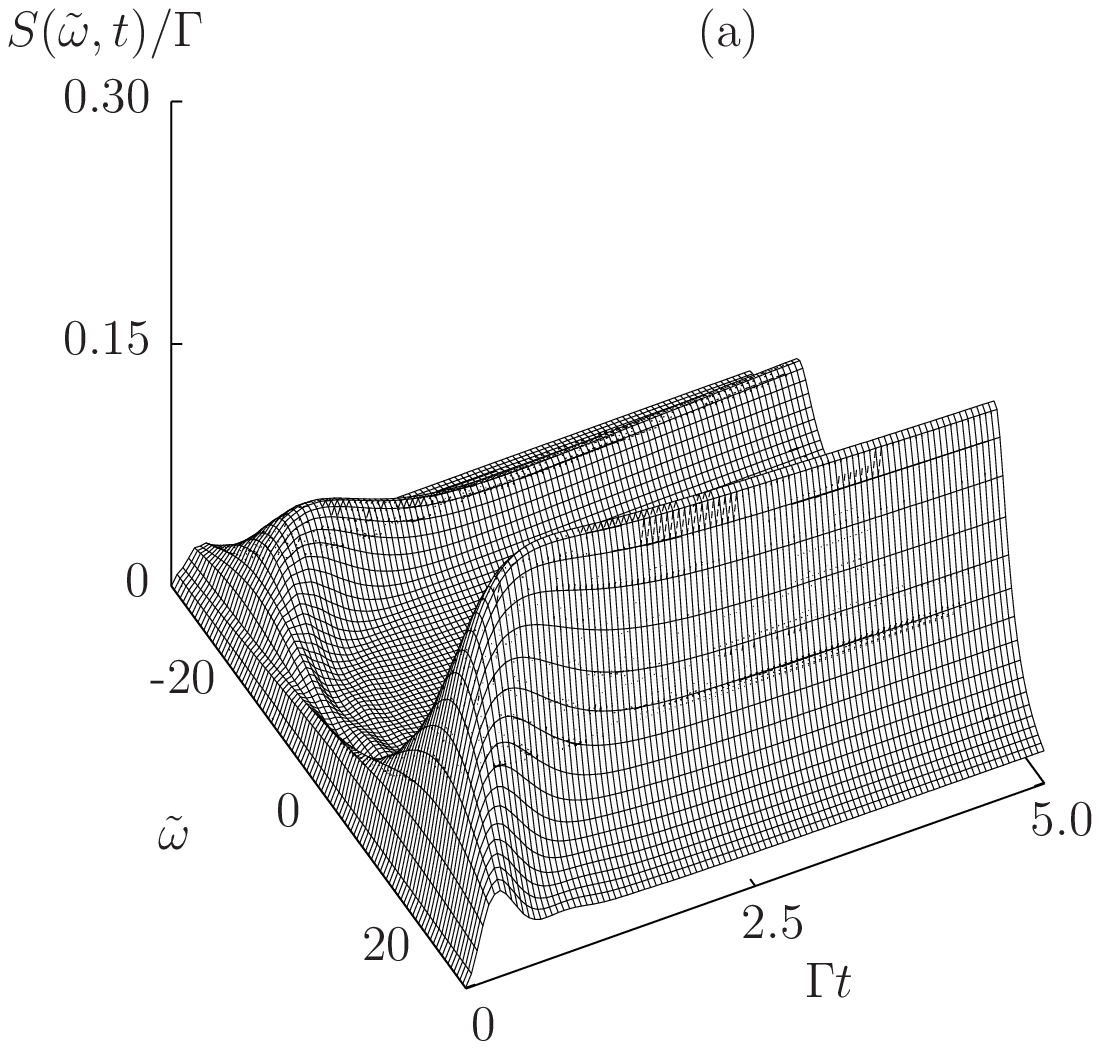}}
\resizebox*{7cm}{!}{\includegraphics{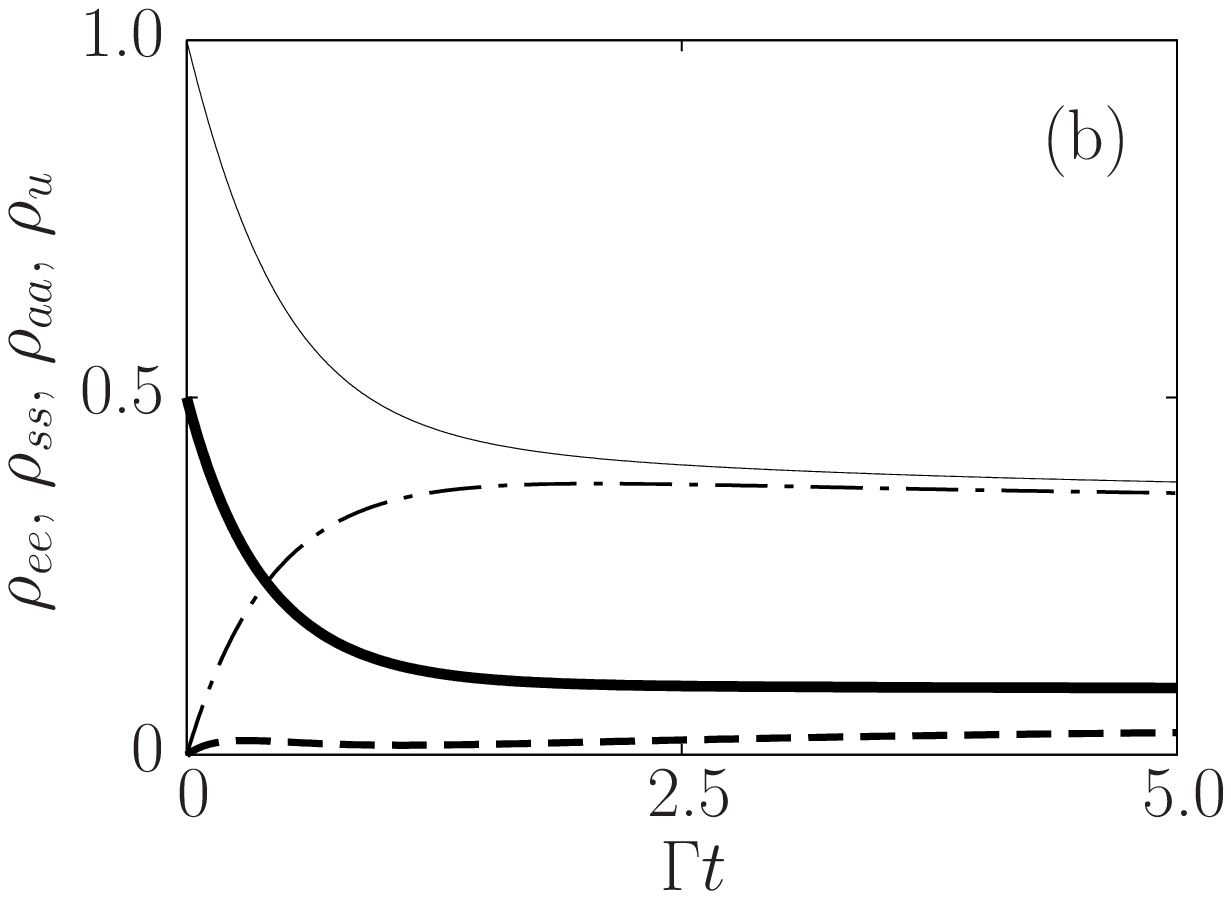}}
\caption{(a) The transient spectrum $S(\tilde{\omega},t)$ as a function of the normalized frequency~$\tilde{\omega} = (\omega -\omega_{0})/\Gamma$ and time $\Gamma t$ for the non-Dicke model with $\Gamma_{d}=2\Gamma$, $\Delta\phi=0$, $N=0.5$, $\vert M\vert=\sqrt{N (N+1)}$, $\Omega_{12}=20\Gamma$, and $\Gamma_{12} =0.5\Gamma$. The frame (b) illustrates the time evolution of the populations $\rho_{ee}$ (thick line), $\rho_{ss}$ (thick dashed line), and $\rho_{aa}$ (dashed-dotted line) together with the two-photon coherence $\rho_u$ (thin line). The atoms were initially in the Bell state (\ref{e13}).}
\label{fig5}
\end{figure}

For phase independent initial states and/or absence of the squeezed field, the spectrum of the field emitted from a two-atom system is composed of two peaks, a broad peak of the width $\Gamma +\Gamma_{12}$ due to spontaneous emission from the symmetric state and a narrow peak of the width $\Gamma -\Gamma_{12}$ arising from spontaneous emission from the antisymmetric state~\cite{le70,hf85}. In this case, the collective damping imposes quantitative changes to the spectrum by modifying the widths of the spectral lines, but does not lead to spectra which are qualitatively different from those that occur in the absence of the collective damping. Here, we show that the inclusion of an initial phase can lead to transient spectra which are quite distinct to those seen in its absence. 

The transient spectrum can be found analytically by solving the equations of motion~(\ref{e11}) and (\ref{e12}). However, the analytical form of the spectrum is very complicated and difficult to interpret. Therefore, we restrict ourselves here to the graphical illustration of the spectra. 

Figure~\ref{fig5}(a) shows the transient spectrum for $\Delta\phi=0$ and $\Gamma_{12}>0$. The spectrum is composed of two asymmetric peaks, whose initial amplitudes decay exponentially in time and attain their stationary values after a short time $t\approx \Gamma^{-1}$. 
The origin of the peaks and their time evolution is easily explained from the analyses of the populations of the collective states, shown in Fig.~(\ref{fig5})(b). Since $\rho_{ee}(0)=1/2$ and $\rho_{ss}(0)=\rho_{aa}(0)=0$, it is clear that the peaks arise from the $\ket e \rightarrow\ket s$ and $\ket e \rightarrow\ket a$ transitions. However, the spontaneous transitions $\ket e\rightarrow\ket s$ do not lead to population of the state $\ket s$ that remains almost unpopulated for all times. A transfer of the initial population occurs only from $\ket e$ to $\ket a$. Moreover, the transfer of the population to the state $\ket a$ occurs on the time scale much shorter than $t=(\Gamma-\Gamma_{12})^{-1}$, the time scale required for appreciable changes of the population of $\ket a$ to occur, as predicted by the rate equations~(\ref{e15}). This indicates that the state $\ket a$ is populated by stimulated rather than spontaneous emission process. 
\begin{figure}[h]
\resizebox*{8.5cm}{!}{\includegraphics{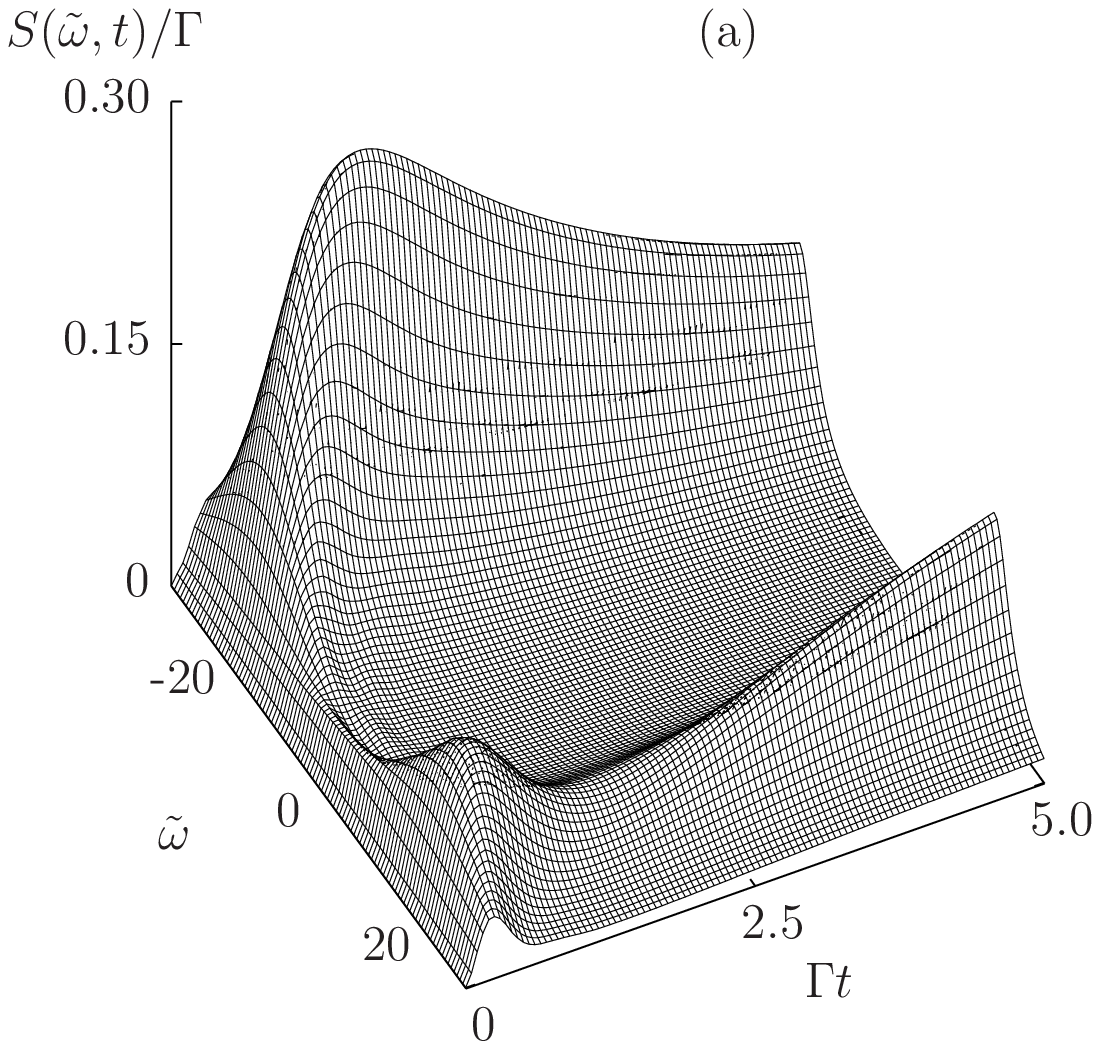}}
\resizebox*{7cm}{!}{\includegraphics{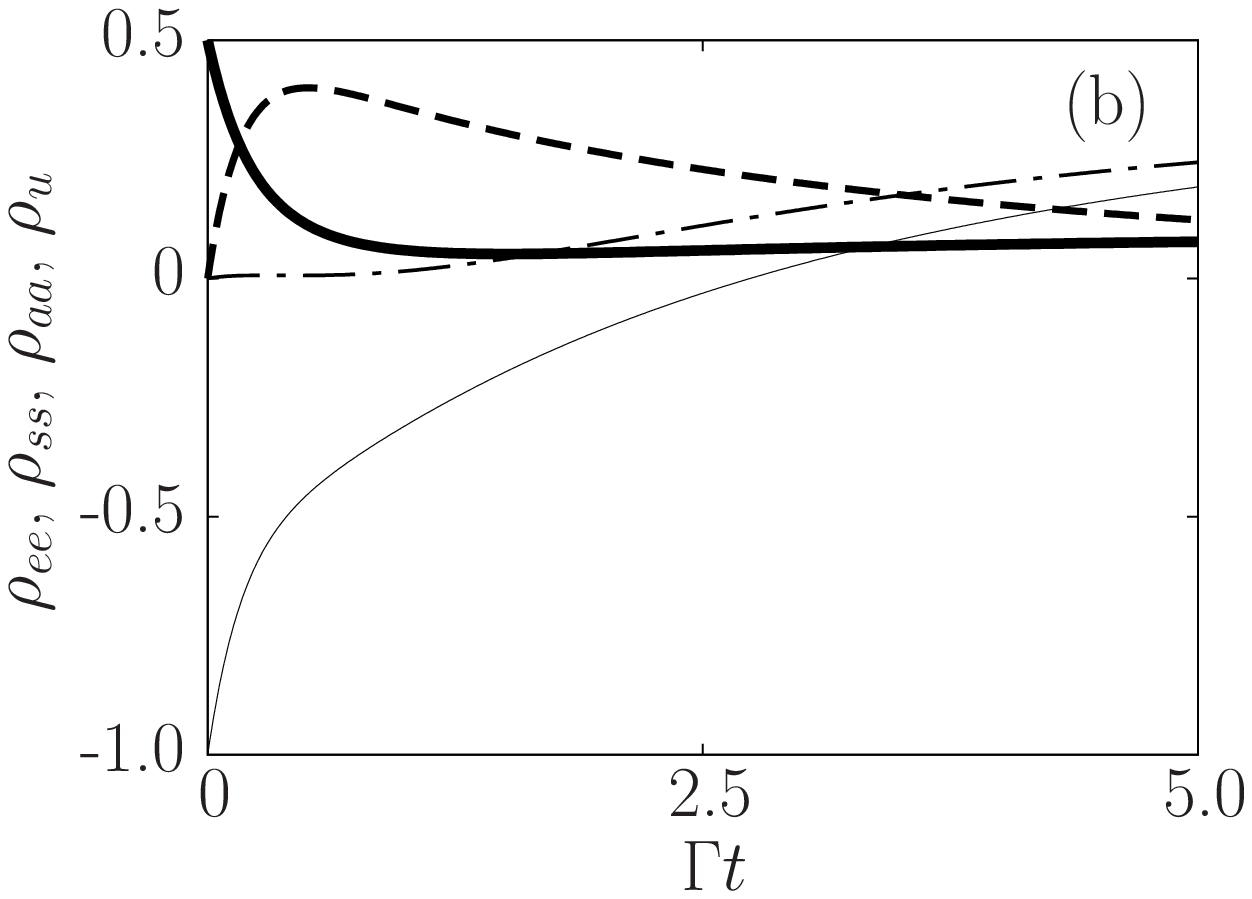}}
\caption{Same as in Fig.~\ref{fig5} but for the relative phase $\Delta\phi=\pi$.}
\label{fig6}
\end{figure}

The situation changes when we switch the relative phase from $0$ to $\pi$. In this case, a hole burning occurs in the higher frequency peak, and simultaneously the amplitude of the lower frequency peak becomes enhanced. This is demonstrated in Fig.~\ref{fig6}, where we plot the spectrum for the same parameters as in Fig.~\ref{fig5}, but with $\Delta\phi =\pi$. The origin of this effect is the same as discussed above for the Dicke model. The explanation again follows from the observation that the destructive interference between the initial two-photon coherence and that induced by the squeezed field leads to spontaneous processes which dominate over the stimulated  processes. As the result, the symmetric state becomes significantly populated on a time scale $t=(\Gamma +\Gamma_{12})^{-1}$, which is the time scale for the population of the symmetric state, predicted by the rate equations. This shows clearly that the phase $\Delta\phi$ changes the transient populations of the collective states of the system. This fact is illustrated in Fig.~\ref{fig6}(b), where we show the time evolution of the populations and the two-photon coherence. Clearly, at the time when the hole burning occurs, the symmetric state is significantly populated and the states $\ket a$ and $\ket e$ are essentially completely depopulated.  
\begin{figure}[h]
\resizebox*{8.5cm}{!}{\includegraphics{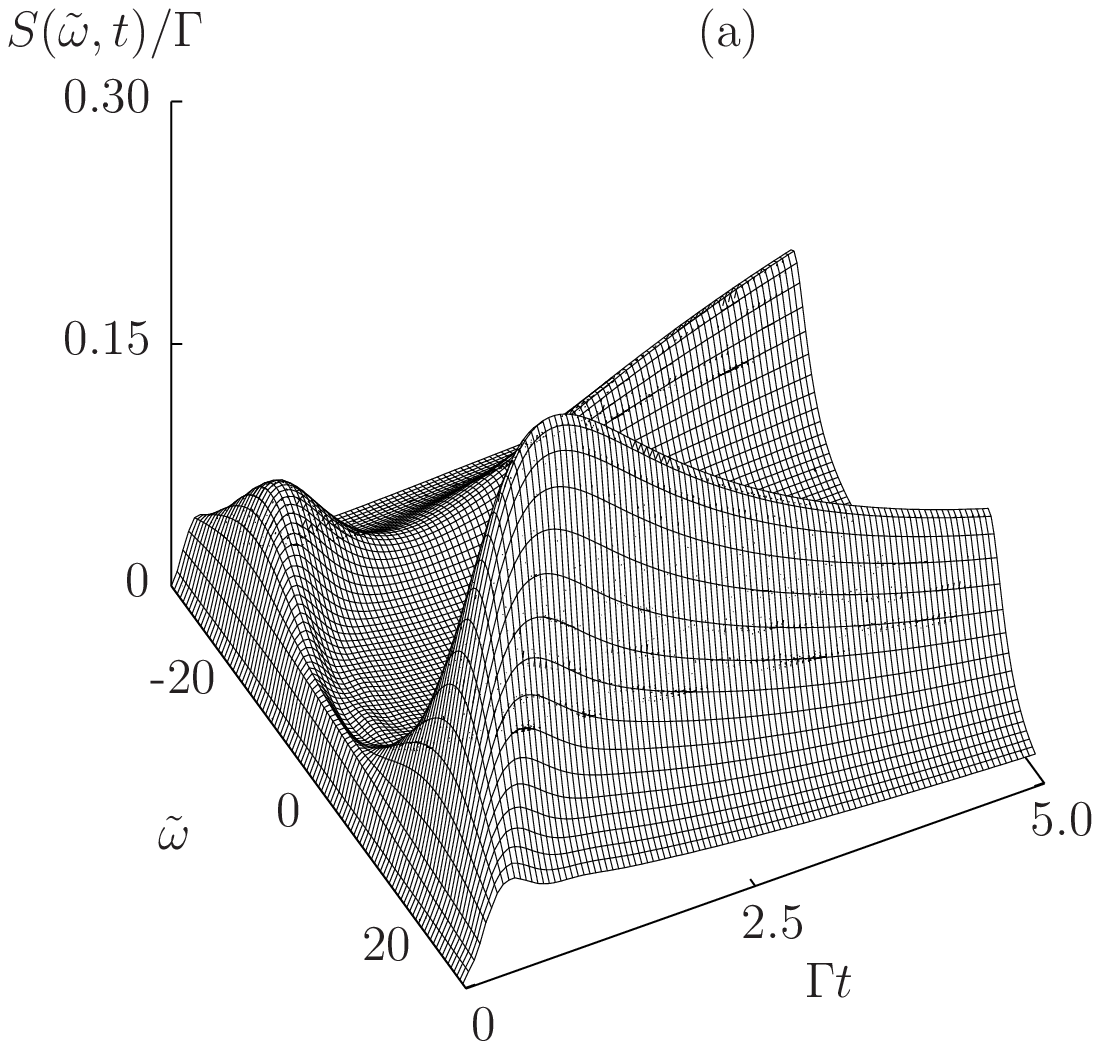}}
\resizebox*{7cm}{!}{\includegraphics{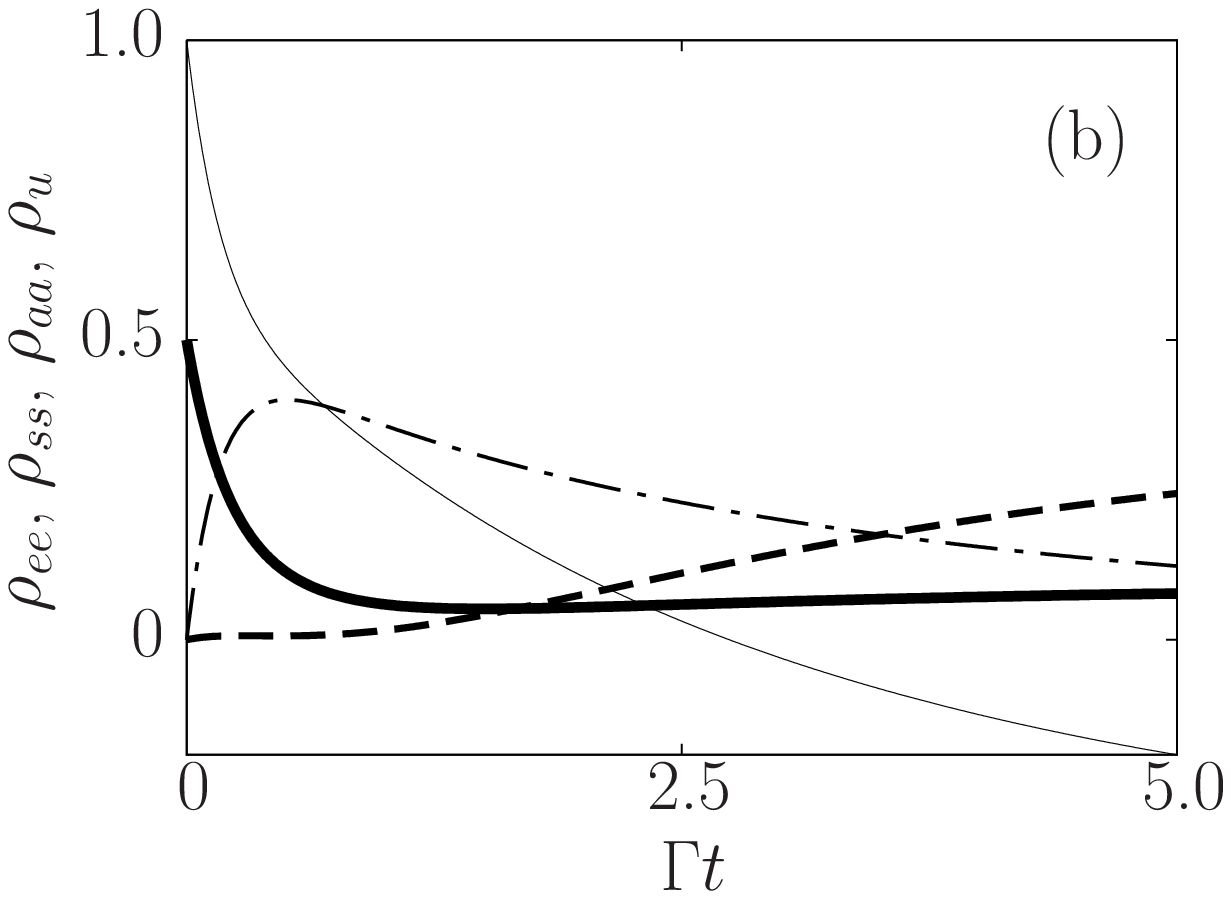}}
\caption{Same as in Fig.~\ref{fig5} but for a negative collective damping rate, $\Gamma_{12}=-0.5\Gamma$.}
\label{fig7}
\end{figure}

It is interesting that hole burning can occur not only by changing the relative phase~$\Delta\phi$ , but also by changing the sign of $\Gamma_{12}$. A change of the sign of $\Gamma_{12}$ from $\Gamma_{12}>0$ to $\Gamma_{12}<0$ is achieved by varying the distance between the atoms. This is shown in Fig.~\ref{fig7}, where we plot the spectrum for the same parameters as in Fig.~\ref{fig5}, but for $\Gamma_{12}=-0.5\Gamma$. We see that the sign of the collective damping parameter, which depends on the phase difference between the oscillating atomic dipoles, has a strong impact on the shape of the spectrum. A change $\Gamma_{12}\rightarrow -\Gamma_{12}$ leads to hole burning in the lower frequency peak. Thus, the effect of negative $\Gamma_{12}$ is seen to be the same as the relative phase $\Delta\phi =\pi$. This is a surprising result as one could expect that a change of the sign of $\Gamma_{12}$ would result in a spectrum similar to that as shown in Fig.~\ref{fig5}, but with the peaks exchanged in position. The changes in the spectral distribution indicate that the phase of the squeezed field is locked to the phase difference between the oscillating dipoles. We may conclude that the collective damping acts like an artificial tweezer that rotates the phase of the squeezed field. To say this another way: The squeezed field keeps memory of the relative phase of the atomic dipole moments. 
\begin{figure}[h]
\resizebox*{8.5cm}{!}{\includegraphics{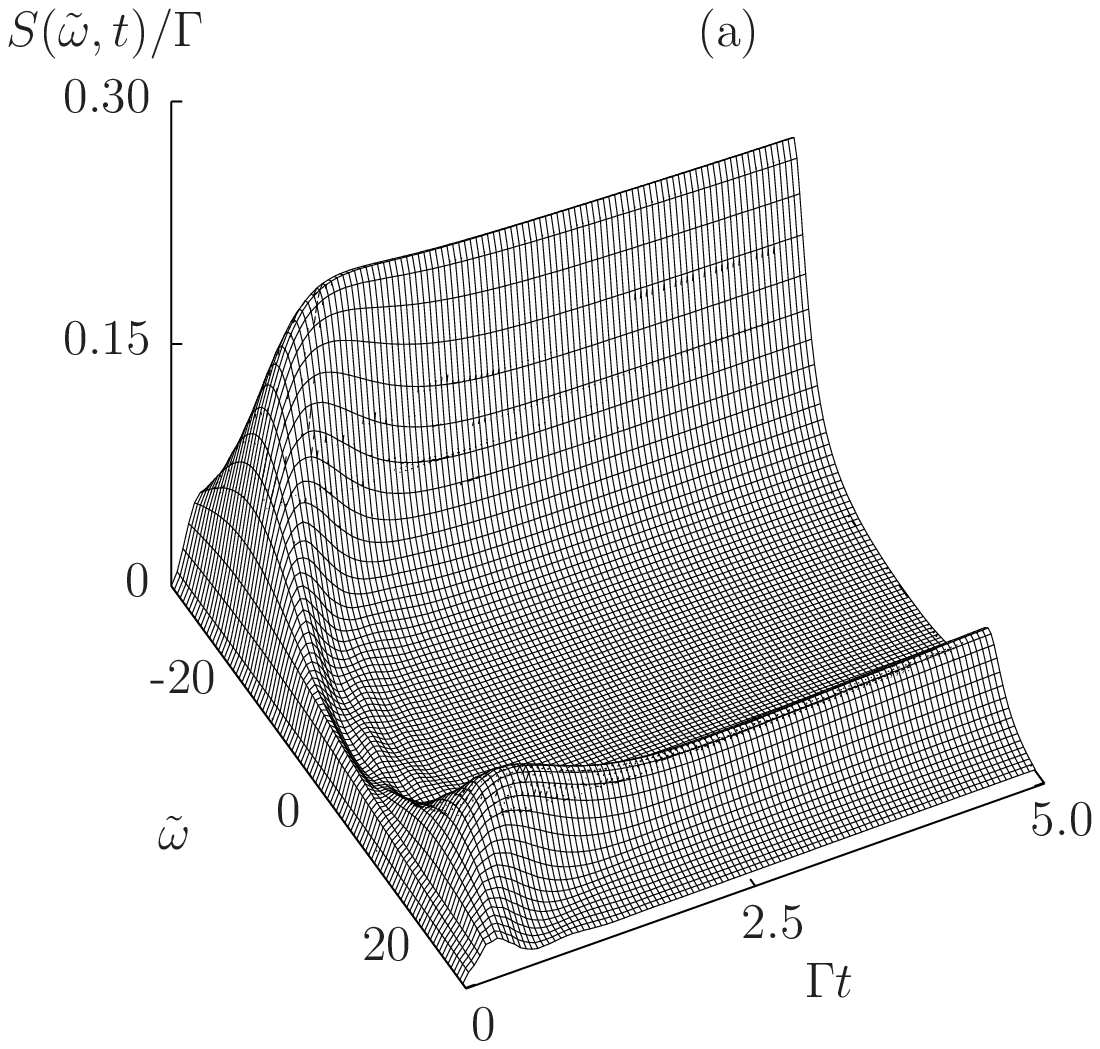}}
\resizebox*{7cm}{!}{\includegraphics{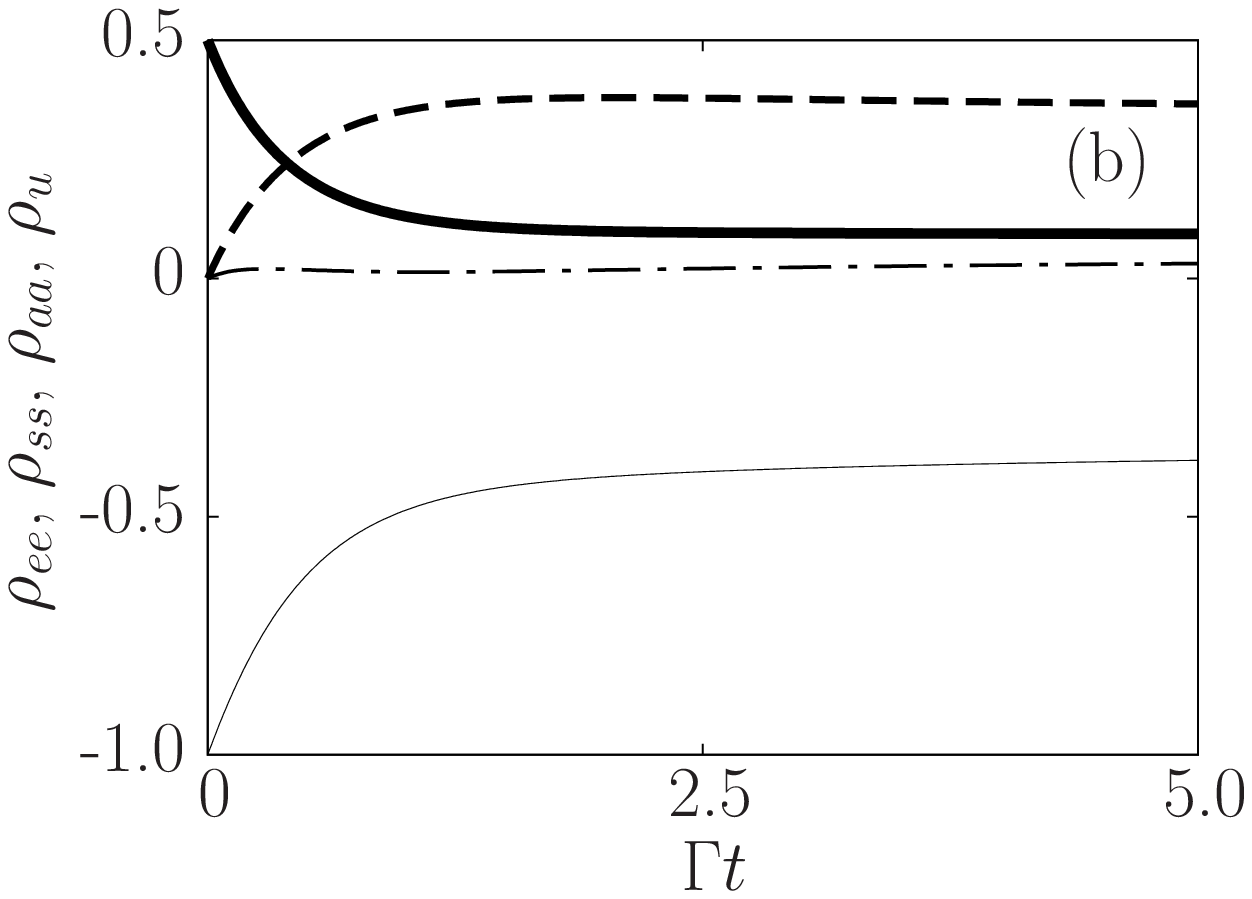}}
\caption{Same as in Fig.~\ref{fig5} but for $\Delta\phi=\pi$ and $\Gamma_{12} =-0.5\Gamma$.}
\label{fig8}
\end{figure}

It is not difficult to see that $\Gamma_{12}$ may act as a tweezer of the squeezing phase. This fact is graphically shown in Fig.~\ref{fig8}, where we plot the transient spectrum for the same parameters as in Fig.~\ref{fig7}, but for $\Delta\phi =\pi$. When $\Delta\phi$ is switched from zero to $\pi$, the spectrum reverts to a profile similar to that shown in Fig.~\ref{fig5}(a) that the spectrum is composed of two asymmetric peaks.

The role of $\Gamma_{12}$ as a phase shifter is also seen from the solution for the two-photon coherence~$\rho_{eg}$. It is not difficult to show from Eq.~(\ref{e15}) that the long time $(t\rightarrow \infty)$ solution for the two-photon coherence $\rho_{eg}$ is of the form
\begin{eqnarray}
\rho_{eg} = \frac{n^{3}|a||M|{\rm e}^{i(\phi_{s}+\Phi_{a})}}{n^{2}\left[n^{4}+4|M|^{2}\left(|a|^{2}-n^{2}\right)\right]} ,\label{e23}
\end{eqnarray}
where $\Phi_{a}$ is a phase angle with only two discrete values of $\Phi_{a} = 0$ for $\Gamma_{12} >0$ and $\Phi_{a}=\pi$ for $\Gamma_{12}<0$. It is clear that a change $\Gamma_{12}\rightarrow -\Gamma_{12}$ leads to a shift of the squeezing phase by $\pi$. It is also interesting to note that the magnitude of the coherence is proportional to $\Gamma_{12}|M|$, which indicates that the collective damping $\Gamma_{12}$ can be regarded as an efficiency of the transfer of the two-photon correlations $|M|$ from the squeezed field to the atoms. 
\begin{figure}[h]
\resizebox*{8cm}{!}{\includegraphics{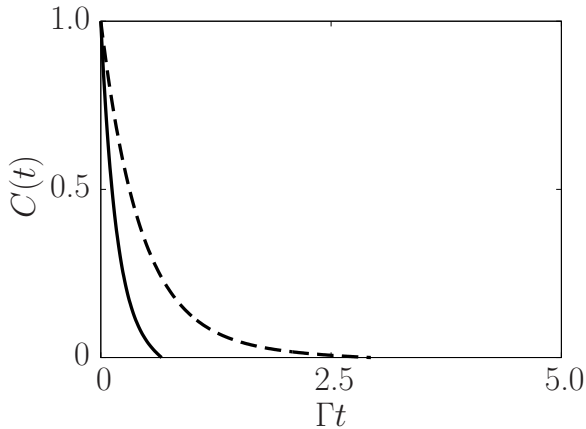}}
\caption{Time evolution of the concurrence for the non-Dicke model coupled to a broadband squeezed vacuum field with $N=0.5$, $\vert M\vert=\sqrt{N (N+1)}$, $\Delta\phi=0$, and different signs of $\Gamma_{12}$: $\Gamma_{12} =-0.5\Gamma$ (solid line), $\Gamma_{12}=0.5\Gamma$ (dashed line).}
\label{fig9}
\end{figure} 

Finally, we compare the transient spectrum with the transient properties of the concurrence. 
Figure~\ref{fig9} shows the concurrence $C(t)$ as a function of time for the same parameters as in Figs.~\ref{fig5} and \ref{fig7}. The decay time of the initial entanglement depends on $\Delta\phi$. We observe that similar to the Dicke model, the initial entanglement collapses, i.e. disappears over a finite time but, what is surprising, it never revives. The lack of recurrence of entanglement in the non-Dicke system is due to the fact that the presence of the antisymmetric state increases the threshold for entanglement in $C_{1}(t)$ from $\rho_{ss}(t)$ to $\rho_{ss}(t)+\rho_{aa}(t)$, as seen from Eq.~(\ref{4.12}), and decreases the positive term in $C_{2}(t)$ from~$\rho_{ss}(t)$ to $|\rho_{ss}(t)-\rho_{aa}(t)|$, as seen from Eq.~(\ref{4.13}). The enhancement of the threshold in~$C_{1}(t)$, it turns out, is sufficient to wipe out any entanglement. Physically, the lack of the entanglement revival is the result of an imperfect  transfer of the two-photon correlations from the squeezed field to the atoms. It is easy to see from Eqs.~(\ref{e15}) and~(\ref{e23}) that the two-photon coherence $\rho_{eg}$ is proportional to $a=\Gamma_{12}/\Gamma$ that is smaller than one $(a<1)$ for the non-Dicke model.

\section{Conclusion}\label{sec6}

We have examined  the transient spectrum of the radiation field emitted by a system of two atoms that interact with a broadband squeezed vacuum field and initially prepared in a maximally entangled Bell state. The initial entangled state provides a reference phase for the phase dependent squeezed vacuum field. We have considered the Dicke and non-Dicke models. The non-Dicke model explicitly includes a phase difference between the atomic dipole moments. Our calculations have demonstrated the qualitatively new behavior observable in the transient spectrum of a collective system that evolves from an initial entangled state. It has been found that the effect of the relative phase suppresses one of the two spectral peaks normally expected in transient spectra and enhances the other. The suppression is seen as a hole burning in the spectral line. The physical origin of the hole burning has been explained in terms of the destructive interference between the initial and the squeezed field induced two-photon coherence. The relation between the phases determines whether the population is transferred to the two-excitation energy state or to a single-excitation collective state. We have also made a comparison of the transient behavior of the spectrum with the time evolution of the initial entanglement and found that the hole burning can be  interpreted as a manifestation of the phenomenon of entanglement sudden death. In addition, we have found that in the case of the non-Dicke model, the collective damping rate that can be positive or negative, introduces a phase shift of the phase of the squeezed field. In all cases we have considered, we have found that a change of the sign of the collective damping results in the hole burning in one of the spectral lines, and the suppression of a given peak  determines the state that is selectively populated.

\section*{Acknowledgments}

This research was supported by the Australian Research Council Discovery program and KACST.

\end{document}